\documentclass{aa}
\usepackage{graphicx}
\usepackage{txfonts}
\usepackage{natbib}
\bibpunct{}{}{;}{a}{}{,}

\begin{document}

\title{A cryogenic waveplate rotator \\ for polarimetry at mm and sub-mm wavelengths}

\author{M. Salatino\inst{1,2}, P. de Bernardis\inst{1,2}, S. Masi\inst{1,2}}

\institute{ Dipartimento di Fisica, Universit\`{a} di Roma ``La
Sapienza", Roma, Italy
    \and INFN Sezione di Roma 1, Roma, Italy}

\offprints{maria.salatino@roma1.infn.it}
\date{Submitted: Jun. 28, 2010; Revised, Dec.27, 2010, Accepted: XXX}

{
    \abstract
{{\bf \sl Context:} Mm and sub-mm waves polarimetry is the new
frontier of research in Cosmic Microwave Background and
Interstellar Dust studies. Polarimeters working in the IR to MM
range need to be operated at cryogenic temperatures, to limit the
systematic effects related to the emission of the polarization
analyzer.

{{\bf \sl Aims:}  In this paper we study the effect of the
temperature of the different components of a waveplate polarimeter,
and describe a system able to rotate, in a completely automated way,
a birefringent crystal at 4$\,$K.}

{{\bf \sl Methods:} We simulate the main systematic effects related
to the temperature and non-ideality of the optical components in a
Stokes polarimeter. To limit these effects, a cryogenic
implementation of the polarimeter is mandatory. In our system, the
rotation produced by a step motor, running at room temperature, is
transmitted down to cryogenic temperatures by means of a long shaft
and gears running on custom cryogenic bearings.

}

{{\bf \sl Results:} Our system is able to rotate, in a completely
automated way, a birefringent crystal at 4$\,$K, dissipating only
a few mW in the cold environment. A readout system based on
optical fibers allows to control the rotation of the crystal to
better than 0.1$^{\circ}$.}

{{\bf \sl Conclusions:} This device fulfills the stringent
requirements for operation in cryogenic space experiments, like
the forthcoming PILOT, BOOMERanG and LSPE.}


 }

\authorrunning{Salatino \emph{et al.}}
\titlerunning{Cryogenic waveplate rotator}
\maketitle

\section{Introduction}

Diffuse dust in our Galaxy is heated by the interstellar radiation field at temperatures ranging from 10$\,$K to 100$\,$K,
de\-pen\-ding on both the dimensions of the grains and the radiative environment. Their emission is thus in the sub-mm/FIR range,
and is partially polarized, due to the alignment of grains in the Galactic magnetic field (see e.g. Draine \cite{Drai03}).

While FIR emission of interstellar dust has been mapped quite accurately, even at high galactic latitudes, by the IRAS, ISO and
Spitzer surveys, and is investigated in deep detail by the currently operating Herschel observatory, its polarization properties
are almost unknown (see e.g. Hildebrand \& Kirby \cite{Hild04}).

The interest in studying dust polarization is twofold. An accurate
measurement of dust polarization at different wavelengths is
important to better understand the nature and structure of dust
grains and to probe the magnetic field of our Galaxy (see e.g.
Vaillancourt \cite{Vail09}, Hildebrand et al. \cite{Hild00}).

Moreover, polarized dust emission is an important
con\-ta\-mi\-na\-ting foreground in precision measurements of the
polarization of the cosmic microwave background (CMB), the current
ambitious target of CMB measurements. Measuring it at wavelengths
where dust polarization is dominant is mandatory in order to correct
the co\-smo\-lo\-gi\-cal signal at the level required to measure
B-modes (see e.g. Lazarian et al. \cite{Laza09}). Diffuse Galactic
sources act as a foreground, mimicking the cosmological polarized
signal (Hanany and Rosenkranz \cite{Hanany03a}; Ponthieu and Martin
\cite{Ponthieu06}): at frequencies above 100$\,$GHz interstellar
dust is the dominant foreground. Therefore, an accurate
know\-led\-ge of the polarization of these sources is necessary to
perform a precise measurement of the polarized cosmological
si\-gnal. This is made difficult by the small amplitude of the
polarized signal of these sources, and by its strong angular and
wavelength dependence (Tucci et al. \cite{Tucci05}).

Several experiments are planned to measure dust polarization at
high galactic latitudes.

The Planck satellite (Tauber et al. \cite{Taub10}) and in particular the High Frequency Instrument (Lamarre et al. \cite{Lama10})
is performing a shallow whole-sky survey of dust polarization at frequencies up to 345$\,$GHz. Here, the signal due to interstellar
dust at high galactic latitudes is roughly similar to the level of CMB anisotropy (Masi et al. \cite{Masi01, Masi06}, Ponthieu et
al. \cite{Ponthieu05}). Higher frequency surveys are more sensitive to the polarization signal of interstellar dust (see Fig.
\ref{fig:0}).
\begin{figure*}
\centering
\includegraphics[width=15cm]{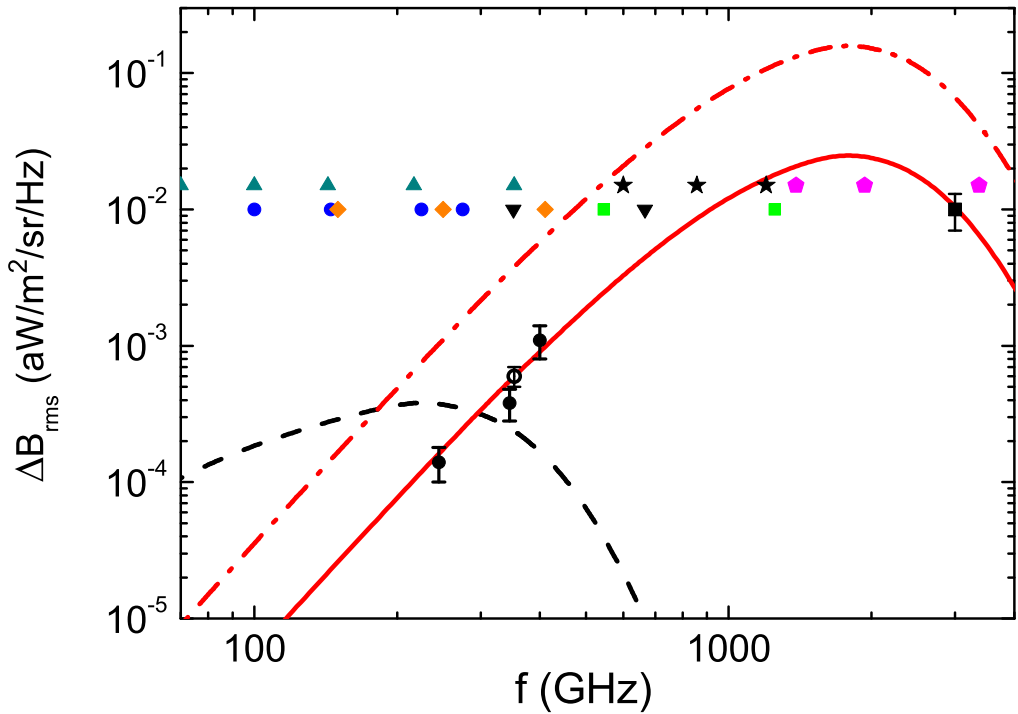}
\caption{ Measurements of rms fluctuations in interstellar dust
emission at Galactic latitudes above 20$^o$ (data points with error
bars) from BOOMERanG (filled circles, Masi et al. \cite{Masi01,
Masi06}) and Archeops (empty circle, Ponthieu et al.
\cite{Ponthieu05}). The continuous line is a thermal spectrum with
$T_d=18\,$K and $\nu^2$ emissivity. The dashed line represents rms
CMB anisotropy for $\Delta T/T = 10^{-5}$. The dash-dotted line
represents the spectrum of the polarized signal from a dust cloud
with brightness of 1 mK$_{RJ}$ at 353$\,$GHz and 10\% polarization
(see Benoit et al. \cite{Benoit04}). The symbols mark the frequency
bands explored by forthcoming polarization measurements from several
experiments: Planck (up triangles), PILOT (squares), BLAST-Pol
(stars), SPIDER (circles), EBEX (diamonds), SCUBA2 (down triangles),
SOFIA (upgraded HAWC: pentagons). } \label{fig:0}
\end{figure*}

PILOT is a stratospheric balloon borne experiment for the
measurement of the polarization of the continuum emission in the
diffuse interstellar medium at frequencies around 545 and
1250$\,$GHz, with a resolution of $3.29'$ and $1.44'$,
respectively (Bernard et al. \cite{Bernard07}).  The experiment is
optimized to perform surveys of the polarized dust emission along
the Galactic plane but also at high Galactic latitudes, where
previous experiments have provided only upper limits (Benoit et
al. \cite{Benoit04}; Ponthieu et al. \cite{Ponthieu05}). PILOT
will probe the large scale distribution of the galactic magnetic
field and the alignment properties of the dust grains providing
strong constraints for dust models, in particular via the
dependence on frequency of the degree of polarization.

BLAST-pol, the polarization sensitive version of the very
successful BLAST balloon telescope, will also search for polarized
dust emission in the galactic plane in the southern hemisphere
(Marsden et al. \cite{Marsden08}, Fissel et al. \cite{Fissel10}).

SPIDER (Crill et al. \cite{Crill08}) is an  ambitious balloon-borne instrument aimed at the measurement of the polarization of the
CMB by means of large arrays of polarization sensitive detectors. Working in the stratosphere, SPIDER can cover high frequencies
(in particular two bands at 225 GHz and 275 GHz), to monitor polarized emission from interstellar dust where it is not negligible
with respect to the polarized component of the CMB.

EBEX is also a balloon-borne CMB polarimeter, aiming at smaller angular scales and with a dust monitor at 410 GHz (Oxley et al.
\cite{Oxle04}).

There are different instrumental techniques to detect faint
polarized signals.  Polari\-za\-tion-sensitive bolometers (PSB,
Jones et al. \cite{Jones03}) have been used in B03 (Masi et al.
\cite{Masi06}) and Planck (Delabrouille and Kaplan
\cite{Delabrouille01}).

Orthomode transducers have been used in WMAP (Jarosik et al.
\cite{Jarosik03}) and other coherent systems.

A stack of a rotating birefringent crystal followed by a static
polarizer implements a Stokes polarimeter, and has been used in the
BRAIN-pathfinder experiment, where the waveplate o\-pe\-ra\-tes at
ambient temperature (Masi et al. \cite{Masi05}).

In MAXIPOL, instead, a DC motor outside the cryostat rotated, by
means of a fiberglass drive shaft crossing the cryostat shell and
the tertiary mirror, a cryogenically cooled Half Wave Plate (HWP),
mounted in the aperture stop of the optical system (Johnson et al.
\cite{Johnson07}).

EBEX will use a cold achromatic dielectric waveplate (Hanany et
al. \cite{Hanany05}) in the cold optics, rotated continuously by
means of superconducting magnetic bearings (Hanany et al.
\cite{Hanany03b}).

SPIDER will also use a large cold HWP, as the very first optical
component in the system (Bryan et al. \cite{Bryan10a}).

In PolKa a reflective system equivalent to a reflective waveplate is
composed of a rotating polarizer close to a reflector, at room
temperature (Siringo et al. \cite{Siri04}). For a review of
po\-la\-ri\-za\-tion modulation techniques see Ade et al.
(\cite{Ade09}).

In parallel to the development of techniques for polarization
modulation, a vigorous theoretical effort attempts to describe the
behavior of real dielectric HWPs (see e.g. Savini et al.
\cite{Savini09}; Bryan et al. \cite{Bryan10a}; Bryan et al.
\cite{Bryan10b}); moreover very pro\-mi\-sing metal-mesh waveplates
have also been proposed (Pisano et al. \cite{Pisano08}).

In this paper we describe the system we have developed for PILOT,
BOOMERanG, and LSPE (de Bernardis et al. \cite{debe09a}), where
the HWP will rotate in steps, and will be kept at cryogenic
temperatures. The paper is organized as follows: in Sec.$\,$\ref{p
conc} we briefly review the principle of o\-pe\-ra\-tion of a HWP
polarimeter, and we prove the necessity of cooling it down to
cryogenic temperatures (Sec.$\,$\ref{p eff}). After listing
(Sec.$\,$\ref{p req}) the experimental requirements for the
necessary cryogenic waveplate rotator (CWR), we present our
implementation of the system; we describe the hardware structure
(Sec.$\,$\ref{p strat}) and the tests we have performed
(Sec.$\,$\ref{p tests}). We conclude in Sec.$\,$\ref{p concl}
summarizing the main features of our system.

\section{The HWP polarimeter}\label{p conc}

A polarimeter sensitive to linear polarization is composed of a HWP,
rotating at frequency $f_0$, followed by a stationary
po\-la\-ri\-zer. Monochromatic linearly polarized light, passing
through the waveplate, emerges still linearly polarized, but with
its polarization vector revolving at 4$f_0$. Unpolarized radiation
is unaffected by the waveplate. If we place a polarizer between the
waveplate and the detector, only the polarized component of the
incoming radiation is modulated by the rotation of the waveplate.
The amplitude of modulation depends on the polarization of the
radiation: it is maximum (null) for radiation totally (not)
polarized. The signal of interest is encoded at 4$f_0$, far from the
spectral region where 1$/f$ noise is important. Any spurious signal
or systematic effect, at a frequency different from 4$f_0$, is
easily removable. The rotating HWP polarimeter offers a substantial
advantage with respect to a PSB: a single detector measures both the
Stokes parameters of the linear polarization, so that the result is
not affected by drift of uncertainties in the relative calibration
of the different detectors of a PSB.

\begin{figure*}
\centering
\includegraphics[width=12cm]{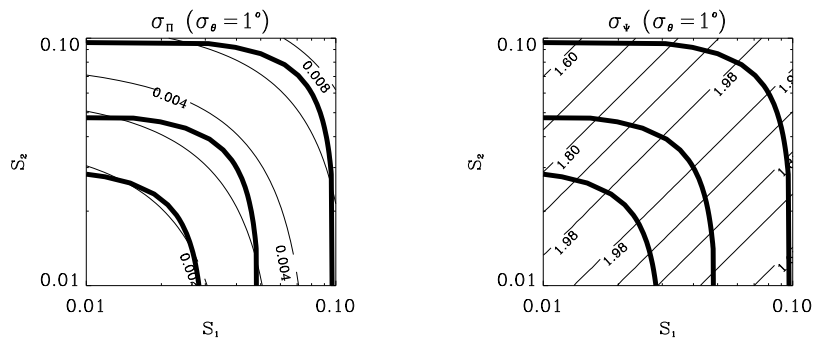}
\caption{Contour levels of the upper limits for the errors in the
polarization degree $\sigma_\Pi$ (left) and in the orientation angle
$\sigma_\psi$ (in degrees, right), caused by a positioning error
$\sigma_\theta$, versus the values of the Stokes parameters $S_1$
and $S_2$ normalized to $S_0$. The figure refers to $N=8$. The thick
contours identify the normalized values of $S_1$ and $S_2$ for which
$\Pi$ is, from bottom left to top right, 3$\%$, 5$\%$, and 10$\%$.}
\label{fig:1}
\end{figure*}

In some cases it is preferable to rotate the HWP in discrete steps.  Our system follows this strategy. The power detected, for a
HWP revolved through $n$ discrete steps, is (Collett \cite{Collett93}):
\begin{equation}\label{e1}
W(n \Delta \theta)=\frac{R}{2}(S_0+S_1\cos{4n \Delta
\theta}+S_2\sin{4n \Delta \theta}),
\end{equation}
where $R$ is the instrument responsivity, $\Delta \theta$ is the step size, and $S_k$ are the Stokes parameters for the incoming
radiation. Eq.$\,$(\ref{e1}) is a truncated Fourier series from which we can recover, given the total number of steps $N$, the
components $S_k$ ($k=0,2$) of the Stokes vector, as:
\begin{eqnarray}\label{e2}
S_0&=&\frac{2}{N}\sum_{n=1}^N W(n \Delta \theta),\nonumber\\
S_1&=&\frac{4}{N}\sum_{n=1}^N W(n \Delta \theta)\cos{4n \Delta
\theta},\nonumber\\ S_2&=&\frac{4}{N}\sum_{n=1}^N W(n \Delta
\theta)\sin{4n \Delta \theta}.
\end{eqnarray}

The drawback of the discrete rotation is enhanced sensitivity to
1$/f$-noise, since the polarized signal is modulated at a frequency
of the order of $f_p=1/T$ where $T$ is the time required to cover
the $N$ steps; given the small value of $f_p$, the noise will be
dominated by instrumental and atmospheric drifts.

From eq.$\,$(\ref{e1}) and (\ref{e2}) we can set upper limits for
the error in the polarization degree, $\Pi=({S_1^2+S_2^2})^{1/2} /
{S_0}$, and for the error in the orientation angle,
$\psi=\frac{1}{2}\arctan{(S_2/S_1)}$, due to inaccuracies in the
position angles of the waveplate. We find

\begin{eqnarray}\label{e3}
\sigma_{\Pi}&\leq&\frac{4\sigma_\theta(|S_1|+|S_2|)\sqrt{4+\Pi^2}}{S_0
\sqrt{N}},\nonumber\\
\sigma_{\psi}&\leq&\frac{4\sigma_{\theta}(|S_1|+|S_2|)}{S_1\sqrt{N}\sqrt{1+{S_2^2}/{S_1^2}}};
\end{eqnarray}

where $\sigma_{\theta}$ is the uncertainty in the position angle
$\theta$ of the waveplate.

To estimate the position accuracy required in a practical case, we
model interstellar dust emission as linearly polarized radiation
with a typical polarization degree $\Pi\sim5\%$ and a typical
specific brightness of about 6$\cdot10^{-16}$W/m$^2$/sr/Hz. Assuming
$\sigma_{\theta}<1^{\circ}$, and using eq.$\,$(\ref{e3}) (i.e.
neglecting other sources of error) with $N=8$, we find an error in
the polarization degree $\sigma_\Pi \lesssim 0.4\%$ (for any
direction of the polarization vector), and an error in the
orientation of the polarization vector $\sigma_\psi \lesssim
2^{\circ}$.

In Fig.$\,$\ref{fig:1} we plot the upper limits for the two errors,
versus the values of the two Stokes parameters of linear
polarization. Increasing the error in the HWP position doesn't
change the shape of the contour levels, given the linear dependence
of the upper limits on this angle (see eq.$\,$(\ref{e3})).

Detector noise is an additional contribution to the error in the
polarization degree and in the orientation of the polarization
vector. This is quantified by the Noise Equivalent Power (NEP) and
by the integration time for each observed pixel ($T$): we have
$\sigma_{det}=$NEP$/\sqrt{T(s)}$, so that:
\begin{eqnarray} \label{e4}
\sigma_{\Pi}&=&\frac{2\sqrt{4+\Pi^2}}{S_0\sqrt{N}}\sqrt{4\sigma_{\theta}^2(|S_1|+|S_2|)^2+\sigma_{det}^2},\nonumber\\
\sigma_{\Psi}&=&\frac{2}{S_1\sqrt{N}\sqrt{1+{S_2^2}/{S_1^2}}}\sqrt{4\sigma_{\theta}^2(|S_1|+|S_2|)^2+\sigma_{det}^2}.
\end{eqnarray}

The specs for the PILOT experiment give a NEP of about
3$\cdot10^{-16}$W/$\sqrt{Hz}$ for the 545$\,$GHz band. In
Fig.$\,$\ref{fig:2} we plot the upper limits on the polarization
degree and on the orientation angle in this case, as a function of
the two Stokes parameters of linear polarization, taking into
account detector noise and integration time as specified above.

For a given degree of polarization, the measurement errors can be
dominated either by detector noise or by positioning errors. We
define the function:
\begin{equation}\label{e5}
g=\frac{4\sigma_{\theta}^2 (|S_1|+|S_2|)^2}{\sigma_{det}^2}.
\end{equation}

When $g<1$ the errors on the polarization degree and the orientation
angles are do\-mi\-na\-ted by detector noise, and can be reduced
increasing the integration time. In the conditions described above
for PILOT, this happens for $\sigma_{\theta}\lesssim0.1^{\circ}$, as
is e\-vi\-dent from Fig.$\,$\ref{fig:2}. If this positioning
accuracy is achieved, one can fully exploit the sensitivity of the
detectors (Fig.$\,$\ref{fig:3}).

\begin{figure}
\centering
\includegraphics[width=8cm]{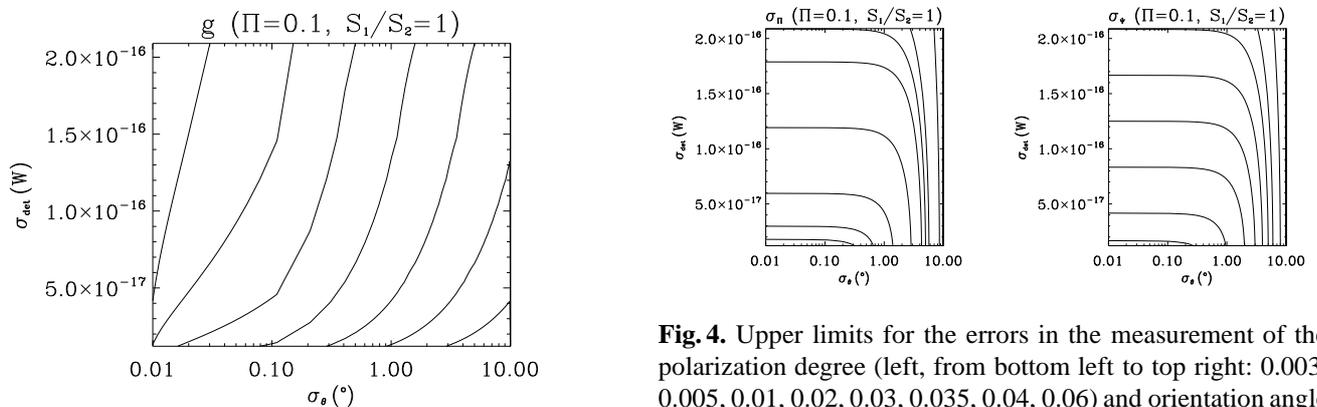}
\caption{Contour levels (from left to right: 1e-4, 1e-3, 1e-2, 1e-1,
1, 10 and 100) for the $g$ function defined in the text,
eq.$\,$(\ref{e5}), versus detector noise, $\sigma_{det}$, and
inaccuracy in the position angle of the waveplate,
$\sigma_{\theta}$. For $g<1$ the errors in the polarization
parameters are dominated by detector noise.} \label{fig:2}
\end{figure}

\begin{figure}
\includegraphics[width=9cm]{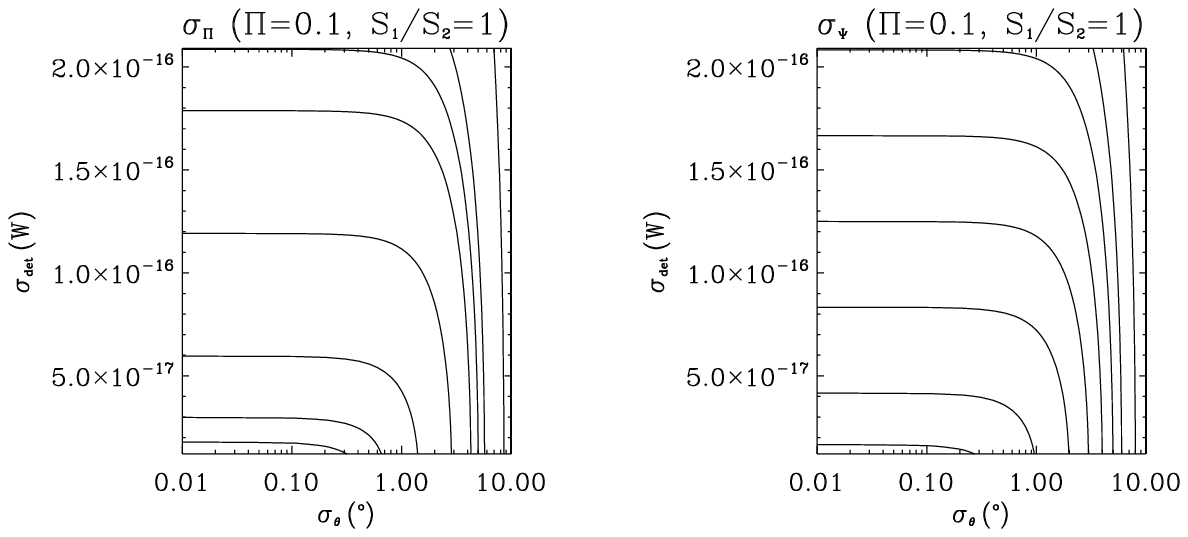}
\caption{Upper limits for the errors in the measurement of the
polarization degree (left, from bottom left to top right: 0.003,
0.005, 0.01, 0.02, 0.03, 0.035, 0.04, 0.06) and orientation angle
(in degrees, right, from bottom left to top right: 0.8, 2, 4, 6, 8,
10, 12, 16), versus detector noise ($\sigma_{det}$) and inaccuracy
in the position angle of the waveplate ($\sigma_{\theta}$).}
\label{fig:3}
\end{figure}

\section{The necessity of cryogenic temperatures}\label{p eff}

\begin{figure*}
\centering
 \includegraphics[width=16cm]{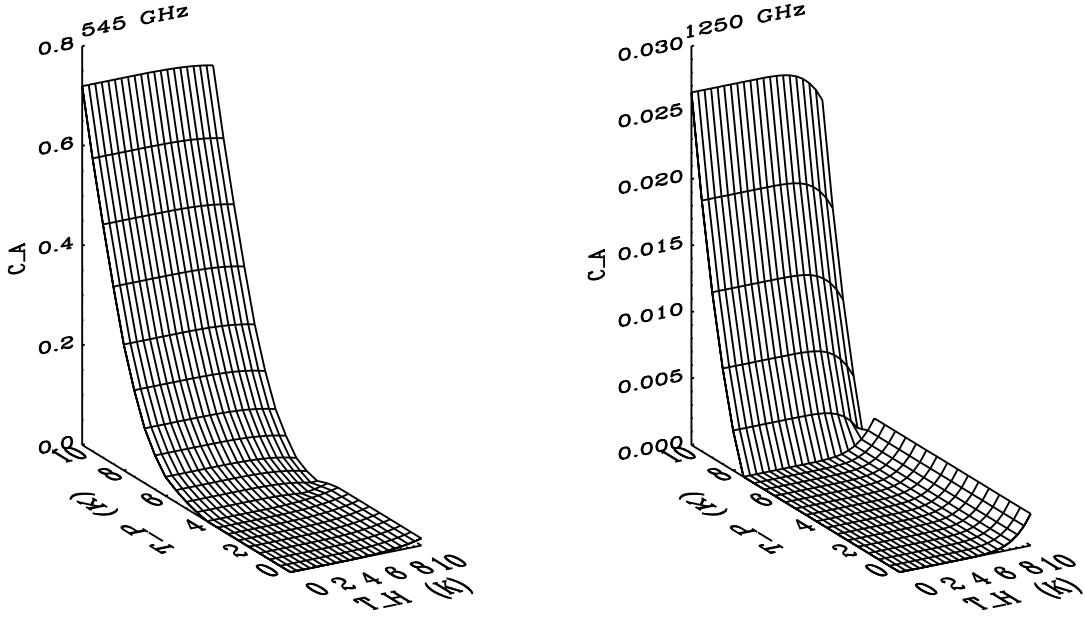} 
 \caption{Plot of the non-ideality
parameter $C_A$ defined in the text, versus the temperature of the
waveplate, $T_H$ and the temperature of the polarizer, $T_P$, and
for two operation frequencies. The non-ideality parameter must be
$<<$1 for a good performance of the polarimeter.} \label{fig:4}
\end{figure*}

Eq.$\,$(\ref{e1}) is valid if the HWP acts as an ideal
phase-shifter with negligible absorption and emission (ideal
waveplate). The detected power is in general a sine wave with
amplitude $A=max(W(n \Delta \theta))-min(W(n \Delta \theta))$, and
offset $O=\frac{1}{N}\sum_{n=1}^NW(n \Delta \theta)$. Non ideal
behavior of the HWP, polarizer, and detector, change the amplitude
and the offset of the detected power, and can introduce spurious
$\cos(2 n \Delta \theta)$ and $\sin(2 n \Delta \theta)$
components.

We have studied how these non-idealities depend on the
o\-pe\-ra\-ting temperature of the polarimeter's components. We
define two non-ideality parameters:
$C_A=|\frac{A_{real}}{A_{ideal}}-1|$ and
$C_O=|\frac{O_{real}}{O_{ideal}}-1|$, where $A_{real\,(ideal)}$ is
the amplitude of the real (ideal) signal and $O_{real\,(ideal)}$ the
offset.

$C_A$ is related to the modulation efficiency of the polarimeter, and compares the actual excursion of the mo\-du\-la\-ted signal
to the excursion in the ideal case (the lossless and efficient polarimeter described by eq.(1)). $C_A \ll 1$ indicates an
instrument close to ideal, thus maximizing the S/N of the measurement.

$C_O$ is related to the offset of the modulated signal, which should be minimized for an optimal use of the dynamic range of the
instrument. A value of $C_O \ll 1$ indicates an offset close to half of the dynamic range, i.e. the ideal case.

For our simulation example we have considered the measurement of polarization of diffuse interstellar dust, which we have modeled
with a temperature of 18$\,$K and emissivity proportional to $\nu^2$, normalized to the Archeops data at 353$\,$GHz (i.e. we used
the spectrum plotted in Fig.$\,$\ref{fig:0}); it is horizontally polarized, with a degree of polarization of $10\,\%$. We have
considered the two wavelengths $240\,\mu$m and $550\,\mu$m, with finite bandwidths $\Delta \lambda/\lambda=41\%$ and 33$\%$,
respectively. We have considered only normal incidence on the HWP. The spectral dependence of the extraordinary and ordinary
refraction indices of the HWP, and of the corresponding absorption coefficients, is given by Savini et al. (\cite{Savini06}). We
have assumed a linear decrease of the absorption coefficient with the temperature of the HWP. Small differences in the absorption
coefficients of the HWP, at a level of about $10^{-3}$, create a modulated polarized emission from the crystal itself. In a typical
Stokes polarimeter (like e.g. MAXIPOL and PILOT) the HWP is followed by a polarizer, tilted of 45$^{\circ}$ with respect to the
normal of the HWP surface. In this way two independent arrays, one detecting radiation transmitted by the polarizer, the other
detecting reflected radiation, can use the same focal plane of the telescope, thus observing the same area of the sky. At these
frequencies, metal wire grid polarizers are close to ideal. In our model we assume transmission coefficients $p_x=0.99$, $p_y=0.01$
and emissivity $0.01$. The radiation is then detected by a bolometer, with emissivity $0.5$, cooled at 0.3$\,$K. Due to this very
low temperature, we have neglected its emission.

The signal detected by the polarimeter, $W_{det}$, is given by the sum of the dust emission processed by the rotating HWP and the
polarizer, the HWP emission processed by the polarizer and the polarizer emission reflected back by the HWP:
\begin{equation}\label{e6}
W_{det}=R(M_{PT}M_{HT}(\theta)S_d+M_{PT}S_{HE}(\theta)+M_{PT}M_{HR}(\theta)S_{PE}).
\end{equation}
In eq.$\,$(\ref{e6}) $M_{PT}$ is the Mueller matrix for the transmission of a polarizer, $M_{HT(R)}(\theta)$ is for the
transmission (reflection) of a HWP, $S_{P(H)E}$ is the Stokes vector for the emission of a polarizer (HWP), and $S_d$ the Stokes
vector for the emission of the dust.

With respect to the ideal case, a real HWP modifies the
a\-stro\-phy\-si\-cal polarized signal reducing the amplitude and
the of\-fset of the transmitted signal. In fact, in our example the
amplitude of the signal transmitted by a real HWP, $A^S_{real}$,
with respect to the ideal one, $A^S_{ideal}$, at 545$\,$GHz is
$A^S_{real}/A^S_{ideal}|_{545}=0.93$; and, similarly, for the
offset: $O^S_{real}/O^S_{ideal}|_{545}=0.93$. The reduction in both
the amplitude and the offset increases at high frequencies; at
1250$\,$GHz we find, in fact: $A^S_{real}/A^S_{ideal}|_{1250}=0.88$
and $O^S_{real}/O^S_{ideal}|_{1250}=0.85$.

Our simulation for the amplitude of the signal (the offset is removed through a low pass filter) clearly shows that to increase the
efficiency of the polarimeter it is necessary to cool down to cryogenic tem\-pe\-ra\-tu\-res both the polarizer and the HWP
(Fig.$\,$\ref{fig:4}). From eq.$\,$(\ref{e6}) the radiation emitted by the polarizer and reflected back by the HWP produces a
$\cos{4 n \Delta \theta}$-si\-gnal with the same phase as the astrophysical source, when the wires of the polarizer are aligned to
the polarization vector of the signal (Salatino \& de Bernardis, \cite{Salatino10}). Depending on the in-band spectrum of the
incoming signal, this emission can add to the astrophysical signal, mimicking a spurious increase of its amplitude. Rotating the
polarizer by $35.^{\circ}26$, this emission becomes out of phase with respect to the astrophysical signal. Moreover, the
po\-la\-ri\-zed emission of the HWP, which directly crosses the polarizer, produces a signal modulated at $2 n \Delta \theta$ in
the detected signal.

The $C_A$ parameter is dominated by the temperature of the polarizer. Only if the HWP is warmer than about 10$\,$K the polarized
emission from the crystal contaminates the $\cos{4 n \Delta \theta}$ signal. Given its definition, $C_A$ includes all the terms
contributing to the signal, including any $\cos{2n\Delta\theta}$ term. Moreover, the $C_A$ parameter depends also on the $\cos{2
n\Delta \theta}$ component because the non linear behavior of incoherent detectors produces a $\cos{4 n \Delta \theta}$ signal if
the $\cos{2 n \Delta \theta}$ one is too large. This additional signal can be se\-ve\-ral order of magnitudes larger than the dust
emission.

Given the larger emissivity of the HWP with respect to the one of
the polarizer (by at least one order of magnitude) and the low
reflectivity of the HWP itself, the $C_O$ parameter is dominated by
the HWP temperature.

We run our model for a 300K waveplate, and found that the signal produced by the HWP is $10^4$ times larger than the dust signal.
This poses extreme requirements for the sta\-bi\-li\-ty of the polarimeter, which are completely relaxed for a cryogenic waveplate.

Both parameters decrease with frequency. This is because the
refraction indices and the absorption coefficients increase with
frequency, but the emission of interstellar dust at 30$\,$K
increases faster (at least for frequencies $<$1500$\,$GHz): the net
result is a decrease of the fractional disturbance.

The $C_A$ parameter (Fig.$\,$\ref{fig:4}) decreases quickly with
the polarizer temperature, and it is $<$0.1 for both channels when
the wire grid is cooled below 8$\,$K.

The $C_O$ parameter, instead, decreases quickly with the HWP temperature it becomes 9.8$\cdot10^{-4}$ and 3.5$\cdot10^{-2}$ at
6$\,$K, and 6.5$\cdot 10^{-5}$ and 8.5$\cdot10^{-3}$ when cooling down the HWP to 4.5$\,$K (higher and lower frequency
respectively).

We have also checked that our results are not very sensitive to reasonable changes in the parameters of the dust spectrum. For
example, increasing the spectral index from 2 to 3, and reducing the temperature from 18$\,$K to 15$\,$K, the variations of the
$C_A$ and $C_O$ parameters are less than 20$\%$.

\section{Experimental requirements}\label{p req}

The rotation of the polarization plane is the same when the optical
axis of the HWP is at 0$^{\circ}$ and at 45$^{\circ}$. One can
rotate more in order to get angle redundancy and check for
systematic effects. For example one can check if the same signals
seen in the $0^{\circ}-45^{\circ}$ interval are seen again in the
$45^{\circ}-90^{\circ}$ interval. Moreover, a good sampling of the
angles is found dividing the $0^{\circ}-45^{\circ}$ interval in four
parts. So it is natural to observe eight positions $\theta_k = k
\times \Delta \theta$ with $k=0..7$ and $\Delta \theta =
11.25^{\circ}$, for a total angular coverage of 78.75$^{\circ}$.

The aim of a polarimeter is to produce a map of the Stokes
parameters over the sky area of interest. So the rotation of the
waveplate has to be complemented by a scan of the sky. In a stepping
polarimeter, the HWP angle is kept still during each scan of the
sky, and stepped to the next angle at the end of the scan.

To maximize the overall observation efficiency, the time taken to
rotate the HWP, from any position to any other one, should be much
less than the duration of each scan. Typically one sky scan can be
about one minute long, so the time required to step the waveplate
angle by $\Delta \theta$ should be $\lesssim10\,$s: this requires
a typical angular velocity of the waveplate of about 0.19$\,$rpm.
The mechanism has to rotate the HWP from position number 1 to
number 8, and then come back to position 1; moreover, this scan
must be repeated for the entire duration of a typical experiment
(a few days). This strategy minimizes microphonic signals (which
are a concern in the case of continuously rotating waveplates,
unless they are levitating on superconducting bearings) and
minimizes the power dissipated on the cryostat during the
rotation.

The accuracy of the position of the HWP should be less than
1$^{\circ}$ to minimize the errors on the polarization degree and
orientation angle (see Sec.$\,$\ref{p conc}). The HWP, placed in the
aperture stop of the optical system, will operate at the temperature
of liquid helium (Sec.$\,$\ref{p eff}). The mechanism has to ensure,
by means of a good thermal contact between its mechanical structure
and the phase shifter, a good thermalization of the birefringent
cry\-stal. Moreover intercepting incoming and outgoing rays from the
HWP itself (vignetting) has to be avoided.

Since the system is cooled by a liquid He bath, one has to limit,
to a maximum of few mW, the power dissipated by the rotation of
the modulator. Moreover, a reliable system working at cryogenic
temperatures is required to control, in an accurate way, the HWP
position.

\section{The cryogenic waveplate rotator}\label{p PCPM}

In Sec.$\,$\ref{p strat} we sketch the hardware concept of our
cryogenic waveplate rotator (CWR hereafter); in Sec.$\,$\ref{p
tests} we present and discuss the tests we have performed, with
particular attention to the measurement of dissipated power produced
when running the CWR at cryogenic temperatures (Sec.$\,$\ref{p
diss}).

\subsection{Mechanical structure}\label{p strat}

An achromatic HWP (Pisano et al. \cite{Pisano06}), 50 mm in
diameter, is mounted on a large hollow
gearwheel\footnote{www.gambinimeccanica.it}, which rotates on a
large thrust ball-bearing. The HWP is surrounded by a 6061-Aluminum
alloy support, and is thermally connected to the Helium liquid
temperature by means of a copper braid. The two halves of the
bearing confine the balls in their grooves and are pressed against
each other through belleville washers, gently compensating any
thermal shrinking and keeping the system on-axis. The rotation of
the hollow gearwheel is obtained by means of a worm-screw. A step
motor running at room tem\-pe\-ra\-tu\-re outside the cryostat
rotates the worm screw by means of a magnetic coupling to cross the
cryostat shell, a long fiberglass tube shaft and a fle\-xi\-ble
joint. The motor is mounted on the top flange of the cryostat to
maximize the distance between the room temperature motor and the
cold section of the cryostat, thus maximizing thermal insulation. A
flexible metal joint connects the vertical shaft to the horizontal
worm-screw acting on the CWR gearwheel. A step motor has been
selected because without brushes can operate continuously in the
stratospheric vacuum environment. The current in the step motor has
been limited to $0.1\,A$ per phase, to reduce the power dissipation
in the motor coils and their self-heating in the absence of
convective cooling. Tests under vacuum show that the motor does not
heat up more than 2$\,$K at a speed of 60 rpm with an
o\-pe\-ra\-tion duty cycle of 5$\%$, while for a 100$\%$ duty cycle
and with a speed of 90 rpm the temperature increase is less than
10$\,$K. At this reduced current the torque from the motor is still
more than enough to reliably rotate the cryogenic mechanism.

A careful choice of the materials and of the mechanical play of the
parts, based on simulated thermal contractions at 4$\,$K, is
mandatory to avoid increased friction and eventually a stall of the
system when cooled at cryogenic temperatures. Smooth rotation of the
worm screw and of the drive shaft is obtained by means of a custom
system which combines thrust ball bearings and belleville washers.
The latter press the bearings keeping the gears on axis during the
rotation and compensate substantial shrinking of the system at
cryogenic temperatures. The presence of a flexible U-joint, in place
of commonly used conical gears, avoids thermal loads produced by the
friction between rotating gears, and mechanical play.

Three pairs of optical fibers sense the rotation angles of the
system.  Each pair consists of a fiber connected to an IR
transmitter at room temperature (820$\,$nm AlGaAs emitters Agilent
Technologies model HFBR-1412), reaching the top surface of the
gearwheel plate, and of a return fiber starting from the bottom
surface of the gearwheel plate and reaching the IR detector at room
temperature (silicon pin photodiodes OSRAM model BPX 61). So the
gearwheel interrupts the three optical fibers pairs. Eight groups of
1 mm dia\-me\-ter holes are drilled through the gearwheel plate in 8
positions, corresponding to the 8 selected rotation angles of the
waveplate. Each pattern reproduces in binary code the index $k$ of
the angle (Fig.$\,$\ref{fig:6}). The operation of the system is
similar to that of an absolute optical encoder: we detect and
identify each integration position analyzing the signals transmitted
through (or blocked by) the gearwheel.

The gap (about 3 mm) in the optical fibers, due to the thickness
of the gearwheel, introduces a significant light loss. This is
recovered by modulating at 1$\,$kHz the IR emitters and recovering
their signal by means of a synchronous demodulator. A block
diagram of the readout electronics is reported in
Fig.$\,$\ref{fig:6}. For gaps smaller than 10$\,$mm (which is our
operating condition) the output signal is highly stable: global
variations range from 0.5$\%$ to 2$\%$ depending on the
amplification gain. A detection test, made with the insertion of a
1$\,$mm diaphragm (the same diameter of the holes in the
gearwheel) between the fibers, leaves unchanged the signal to
noise ratio.

\begin{figure*}
\centering
\includegraphics[width=14cm]{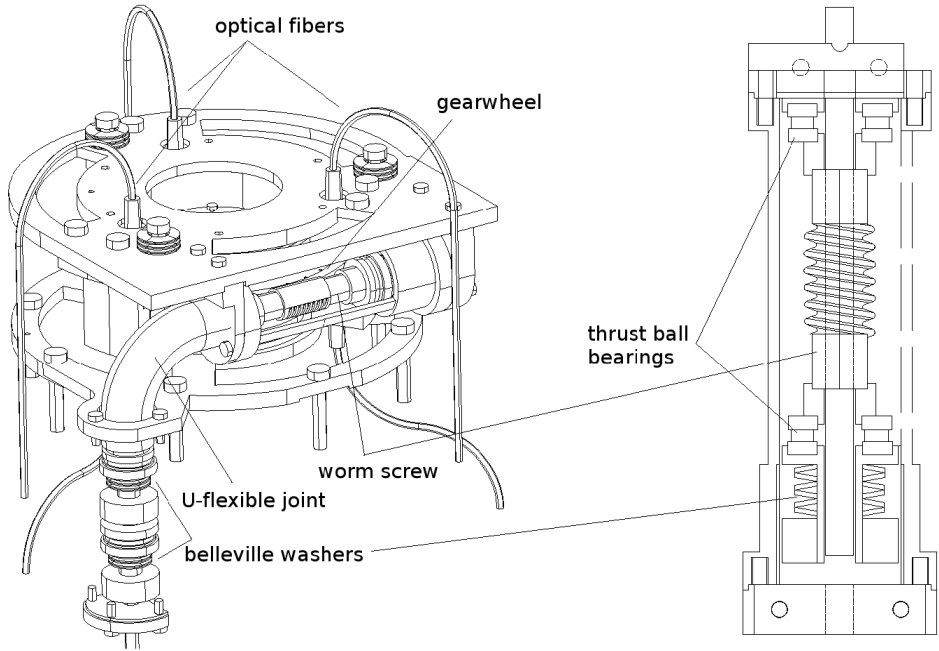} 
\caption{CWR mechanical system (left), and a detail of the custom
cryogenic bearings (right).} \label{fig:6}
\end{figure*}

\begin{figure*}
\centering
\includegraphics[width=12cm]{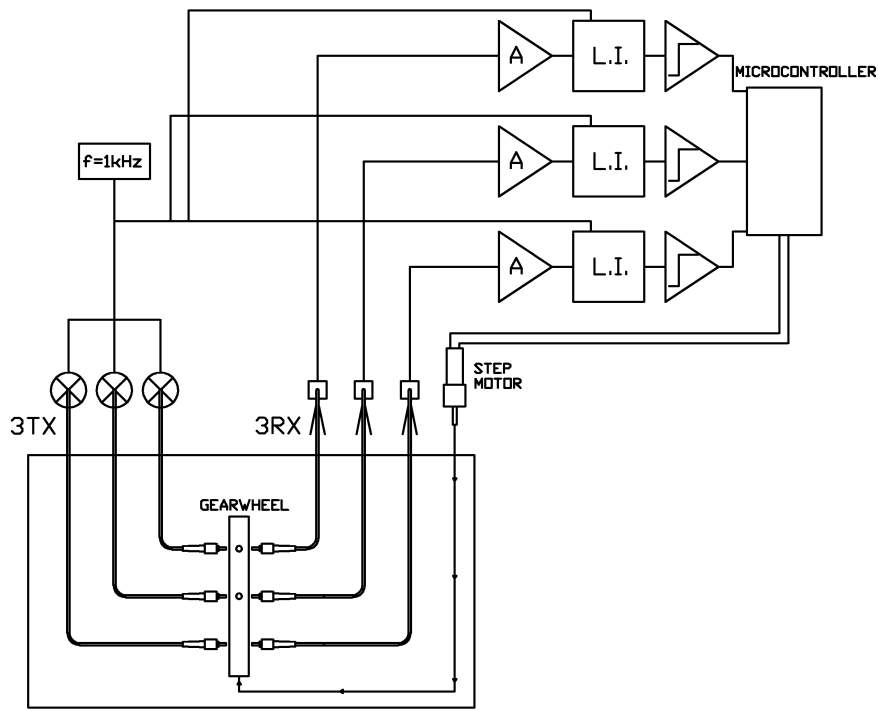} 
\caption{Principle of operation of the CWR electronics. The
modulated light of 3 IR emitters at room temperature (TX) is
transferred via optical fibers to one side of the gearwheel, where
a suitable pattern of holes identifies the 8 integration
positions. Through this pattern of holes three optical fibers,
facing the opposite side of the gearwheel and aligned to the
previous ones, receive the modulated light and transfer it back to
three IR detectors (RX) at room temperature. The signals are
amplified (A) and synchronously demodulated (L.I.), providing
three bits representing the current position angle.} \label{fig:7}
\end{figure*}

\subsection{Qualification Tests}\label{p tests}

We have carried out a number of qualification tests on the key
components of this system.

The CWR, mounted on the cold stage of the photometer, will work
with its axis approximately vertical. During its rotation we need
to avoid shifts of the HWP. We have verified that the shift of the
rotation axis with respect to the aperture stop is limited to
$\lesssim$ 0.13 mm when the CWR is rotated from a vertical axis
configuration to an horizontal one. This shift is completely
acceptable, and has been obtained by means of suitable belleville
washers pressing on the thrust bearing of the gearwheel.

The accuracy and repeatability of the rotation angles of the HWP has
been tested using a laser reflected on a small mirror mounted on the
gearwheel. For all positions, from the histogram of many
measurements we find that the 1-$\sigma$ dispersion  of the position
angle is always well below 1$^{\circ}$ (Tab.$\,$\ref{tab:1})
satisfying the experimental requirements (Sec.$\,$\ref{p req}).

\begin{table}
\caption{Repeatability of the eight HWP position angles. The first
column labels the position number, the second is the mean position
angle, the third the standard deviation of the position angle}
\label{tab:1} \center
\begin{tabular}{lrr}
\hline\noalign{\smallskip}
Position & Nominal angle & Measured standard deviation \\
\noalign{\smallskip}\hline\noalign{\smallskip}
1        & $ 0.00^{\circ}$              & $0.01^{\circ}$ \\
2        & $11.25^{\circ}$              & $0.02^{\circ}$ \\
3        & $22.50^{\circ}$              & $0.01^{\circ}$ \\
4        & $33.75^{\circ}$              & $0.01^{\circ}$ \\
5        & $45.00^{\circ}$              & $0.01^{\circ}$ \\
6        & $56.25^{\circ}$              & $0.01^{\circ}$ \\
7        & $67.50^{\circ}$              & $0.02^{\circ}$ \\
8        & $78.75^{\circ}$              & $0.01^{\circ}$ \\
\noalign{\smallskip}\hline
\end{tabular}
\end{table}

Optical fibers have been tested sunk in liquid nitrogen; in this
con\-fi\-gu\-ra\-tion, the transmitted signal shows an increase of
about 38$\%$. Tests with a He leak-detector have excluded
significant outgassing from the optical fibers and the flexible
U-joint.

The thermal conductivity of the optical fibers has been measured
in the temperature range (4$\div$300)$\,$K. Fiber specimens,
running between the base temperature and a controlled higher
temperature, have been mounted inside a test cryostat. We
estimated the thermal conductivity from the power required to heat
one side of the fibers, keeping the other side at 4K. The
conductivity of the support system and the radiative heat leak
have been properly taken into account. The estimated conductive
thermal load from 4$\,$K to the first thermal shield of the
cryostat, where they will be placed, carried by three pairs of
fibers, 30$\,$cm long, is 0.1$\,$mW.

The fiberglass driveshaft (length 710$\,$mm, inner and outer
diameter 2$\,$mm and 3$\,$mm, respectively), carries a conductive
thermal load, from room temperature to liquid Helium, of 1.0$\,$mW.
So the total static heat load resulting from our system (driveshaft
plus optical fibers) is about 1.1$\,$mW, well within the
experimental requirements (Sec.$\,$\ref{p req}).

\subsection{Power dissipation in dynamic conditions}\label{p diss}

The operation of the CWR at cryogenic temperatures has been tested
inside a la\-bo\-ra\-to\-ry cryostat cooled by a Pulse Tube cooler
(PT). Key temperatures are read by four silicon diode thermometers
mounted on the two stages of the PT, the CWR lid and the top of
the CWR vertical cylinder. A data acquisition/switch unit reads
the thermometers. The optical fibers, 2$\,$m long, cross the
coldest environment and reach the 300$\,$K stage after
thermalization on the intermediate PT stage. A Mylar shield
encloses them avoiding conductive loads due to the contact with
the 300$\,$K shield. A co-axial magnetic coupling transmits the
motor torque through the vacuum shell of the cryostat; a
stainless steel non-magnetic containment barrier allows complete
insulation of the inner magnetic hub from the outer one without
any contact.

The dissipated power is measured  comparing the heating of the
4$\,$K environment, induced by the CWR movement, to the ones
produced by two power wirewound resistors, mounted on the CWR lid.

During this test the intermediate PT stage reached $(68\pm1)\,$K; the low temperature one $(4\pm1)\,$K, the CWR lid $(6\pm1)\,$K
and the CWR cylinder $(6\pm1)\,$K. The torque required to rotate the system was lower than 0.08 N$\cdot$m.

With the CWR thermalized at cryogenic temperature, we have performed a number of scans during which the electronic system read all
the predetermined positions.

We have performed a 50 minutes heat load test, ro\-ta\-ting the waveplate and stopping in each integration position for 30$\,$s.
The residual friction generated during the rotation produced an increase of the tem\-pe\-ra\-tu\-re of the CWR lid ranging from 0.4
to 1.2$\,$K, depending on the rotation speed.

The dissipated power produced by the rotation has been estimated dissipating Joule power in a resistive heater mounted on the CWR,
with the CWR steady. We had to dissipate from 4 to 15 mW, to produce the same temperature increases of the CWR movement.

A change of a fraction of a Kelvin of the HWP tem\-pe\-ra\-tu\-re from a base temperature lower than 6K results in a small change
of the $C_A$ and $C_O$ parameters, which are already small (see Fig.$\,$ \ref{fig:4}). The net result is a negligible effect on the
detected signal.

The total time the system has been operated at low temperature is comparable to the duration of a balloon flight. No wear of the
mechanism has been observed yet. This was expected, since the mechanical stress on the different parts is very small.

\section{Conclusions}\label{p concl}

In this work we have demonstrated the necessity of cooling down to cryogenic tem\-pe\-ra\-tu\-res the optical components of a HWP
polarimeter in order to improve its efficiency and reduce some systematic effects. We have presented the mechanical design and
practical implementation of a cryogenic rotator for a birefringent crystal. The system runs at cryogenic temperature, with an
accuracy in the control of the rotation angle better than 0.1$^{\circ}$. The total heat load on the cryogenic environment is about
1.1$\,$mW (with motor off) and between 4 and 15$\,$mW (when moving). This system meets the requirements of the PILOT stratospheric
balloon experiment (Bernard et al. \cite{Bernard07}), which will study in the near future the polarization of interstellar dust
emission, and is potentially useful for new CMB polarization missions, like LSPE (de Bernardis et al. \cite{debe09a}) BOOMERanG and
B-pol (de Bernardis et al. \cite{debe09b}).

\acknowledgements

We thank A. Schillaci, J.P. Bernard and M. Bouzit for discussions
and suggestions. This work has been supported by the Italian Space
Agency, contracts COFIS 2007-2010 and ``B2K5-Continuazione".

\bibliographystyle{aa}

\begin{thebibliography}{}
\bibitem[2009]{Ade09} Ade, P.A.R., et al. 2009, J. Phys. Conf. Ser., 155, 012006
\bibitem[2004]{Benoit04} Benoit, A., et al. 2004, A$\&$A, 424, 571
\bibitem[2007]{Bernard07} Bernard, J.P., et al. 2007, EAS Publication Series, 23, 189
\bibitem[2010a]{Bryan10a} Bryan, S.A., et al. 2010 [\texttt{ArXiv:astro-ph/1006.3874}]
\bibitem[2010b]{Bryan10b} Bryan, S.A., et al. 2010 [\texttt{ArXiv:astro-ph/1006.3359}]
\bibitem[1993]{Collett93} Collett, E. 1993, Polarized light. Fundamental and applications. (Marcel Dekker, Inc.)
\bibitem[2008]{Crill08} Crill, B.P., et al. 2008, Proc. of SPIE, 7010, 70102P
\bibitem[2009a]{debe09a} de Bernardis, P., et al. 2009, Nuclear Physics B (Proc. Suppl.), 194, 350
\bibitem[2009b]{debe09b} de Bernardis, P., et al. for the B-Pol collaboration 2009, Exp. Astron., 23, 5
\bibitem[2001]{Delabrouille01} Delabrouille, J., $\&$ Kaplan, J. 2001, Proc. of the Pol2001 Astrophysical Polarized Backgrounds conference,
Bologna, Italy
\bibitem[2003]{Drai03} Draine, B.T. 2003, Annual Review of Astronomy and Astrophysics, 41, 241
\bibitem[2010]{Fissel10} Fissel, M.L., et al. 2010, SPIE, 7741, 77410E
\bibitem[2003]{Hanany03a} Hanany, S., $\&$ Rosenkranz, P. 2003, New Astron. Rev., 47, 1159
\bibitem[2003]{Hanany03b} Hanany, S., et al. 2003, IEEE Trans. Appl. Supercond., 13, 2128
\bibitem[2005]{Hanany05} Hanany, S., et al. 2005, Appl. Opt. 44, 4666
\bibitem[2000]{Hild00} Hildebrand, R., et al. 2000, PASP, 112, 1215
\bibitem[2004]{Hild04} Hildebrand, R., $\&$ Kirby, L. 2004, in ASP Conf. Ser. 309,
Astrophysics of Dust, ed. A.N. Witt, G.C. Clayton, $\&$ B.T. Drain,
515
\bibitem[2003]{Jarosik03} Jarosik, N., et al. 2003, ApJ Suppl., 145, 413
\bibitem[2007]{Johnson07} Johnson, N.R., et al. 2007, ApJ, 665, 42
\bibitem[2003]{Jones03} Jones, W., et al. 2003, SPIE, 4855, 227
\bibitem[2010]{Lama10} Lamarre, J.M., et al. 2010, A\&A, 520, A9
\bibitem[2009]{Laza09}Lazarian, A., et al. 2009, in White
Paper to the Cosmology and Fundamental Physics (GCT) Science
Frontiers Panel of the Astro2010 Decadal Survey [\texttt{ArXiv:
astro-ph/0902.4226}]
\bibitem[2008]{Marsden08} Marsden, G., et al. 2008, Proc. of SPIE, 7020, 702002
\bibitem[2001]{Masi01} Masi, S., et al. 2001, ApJL, 553, L93
\bibitem[2005]{Masi05} Masi, S., et al. 2005, EAS Publication Series, 14, 87
\bibitem[2006]{Masi06} Masi, S., et al. 2006, A\&A, 458, 687
\bibitem[2004]{Oxle04} Oxley P., et al. 2004, Proc. SPIE Int.Soc.Opt.Eng. 5543, 320
\bibitem[2006]{Pisano06} Pisano, G., et al. 2006, Appl. Opt., 45, 6982
\bibitem[2008]{Pisano08} Pisano, G., et al. 2008, Appl. Opt., 47, 6251
\bibitem[2005]{Ponthieu05} Ponthieu, N., et al. 2005, A$\&$A, 444, 327
\bibitem[2006]{Ponthieu06} Ponthieu, N., $\&$ Martin, P.G. 2006, Proc. of CMB and Physics of the Early Universe, Ischia, Italy. ed. G. De
Zotti
\bibitem[2010]{Salatino10} Salatino, M., \& de Bernardis, P. 2010, Proc. of the 45th Rencontres de Moriond, La Thuile, Italy, ed. R. Ansari et al.
[\texttt{ArXiv:astro-ph/1006.3225}]
\bibitem[2006]{Savini06} Savini, G., Pisano G., $\&$ Ade, P.A.R. 2006, Appl. Opt., 45, 8907
\bibitem[2009]{Savini09} Savini, G., et al. 2009, Appl. Opt., 48, 2006
\bibitem[2004]{Siri04} Siringo, G., et al. 2004, A\&A, 422, 751
\bibitem[2010]{Taub10} Tauber, J. A., et al. 2010, A\&A, 520, A1
\bibitem[2005]{Tucci05} Tucci, M., et al. 2005, MNRAS, 360, 935
\bibitem[2008]{Vail09} Vaillancourt, J.E. 2008, in ASP Conf. Ser., Astronomical Polarimetry, ed. P. Bastien \& N. Manset [\texttt{ArXiv: astro-ph/0904.1979}]

\end{thebibliography}

\clearpage

\end{document}